# Controllable soliton dragging by dynamical optical lattices

Yaroslav V. Kartashov* and Lluis Torner

*ICFO-Institut de Ciencies Fotoniques, and Department of Signal Theory and Communications, Universitat Politecnica de Catalunya, 08034 Barcelona, Spain*

Demetrios N. Christodoulides

*School of Optics/CREOL, University of Central Florida, 32816-2700, Florida, USA*

We report on the phenomenon of controllable soliton dragging by dynamical optical lattices induced by three imbalanced interfering plane waves. Because of such an imbalance, the transverse momentum of the lattice does not vanish, and thus the dynamical lattice can cause soliton dragging. The dragging rate is shown to depend on the amplitude and on the angle of incidence of the third plane wave making the optical lattice.

The periodic modulation of the linear refractive index of a medium can drastically affect the diffraction properties of light beams. This effect acting in conjunction with nonlinear processes can lead to a specific class of optical solitons – the so-called discrete solitons whose mobility in the transverse direction depends strongly on their power level.[1,2] Arrays of evanescently coupled waveguides are prime examples of such systems where discrete solitons can be directly observed and investigated. The potential of these self-localized states toward practical applications such as all-optical soliton steering and switching has also been addressed in several studies.[3-8]

Quite recently the possibility of creating reconfigurable index profiles has been demonstrated in biased photorefractive crystals. In these latter systems, optical lattices (harmonic refractive index modulations) with adjustable depth and period have been optically induced.[9-11] It is important to note, that so far, all the lattices considered in the literature (see for example Refs. [9-11]), are by nature invariant along the direction of propagation (longitudinal direction). This is because these arrays are established using non-diffracting interference patterns that are polarized in a direction in which the crystal behaves in a quasi-linear fashion. On the other hand, the polarization of the discrete soliton beam is such that it feels not only the periodic potential introduced by the lattice forming beam but also the nonlinearity of the biased photorefractive crystal.[9-10] The possibility of soliton steering and switching in optical lattices has also been analyzed in Refs. [12,13].

In this Letter we investigate the behavior of discrete solitons in dynamic optical lattices that exhibit a finite momentum in the transverse plane. Such lattices can be readily synthesized as long as the sum of the transverse wavevectors of the interfering beams (that optically induce the lattice) is not zero. In this case we show that transverse momentum can be transferred from the lattice to the discrete soliton beam thus leading to a controllable drift. This effect is examined in detail in an imbalanced three-wave lattice interaction where it is shown that the rate of the soliton drift depends on the amplitude and the propagation angle of the third weak plane wave.

Let us consider the propagation of a slit (planar) laser beam along the longitudinal $\xi$ axis in a biased photorefractive crystal. A periodic index modulation is also optically induced in the crystal's transverse $\eta$ direction. The soliton beam and the interfering lattice-creating plane waves are orthogonally polarized to each other so that the nonlinearity affects only the soliton beam.[9-10] Under these conditions, the complex dimensionless field amplitude $q$ evolves according to[9-10]

$$i\frac{\partial q}{\partial \xi} = -\frac{1}{2}\frac{\partial^2 q}{\partial \eta^2} - \frac{Eq}{1+S|q|^2 + R(\eta,\xi)}[S|q|^2 + R(\eta,\xi)]. \tag{1}$$

Here the longitudinal $\xi$ and transverse $\eta$ coordinates are normalized to the diffraction length and the characteristic beam width, the parameter $E$ describes the biasing static field that is applied to the photorefractive crystal, and $S$ represents a saturation parameter. We also assume that the optical lattice is created from three interfering plane waves, i.e., $a\exp(\pm i\alpha\eta)\exp(-i\alpha^2\xi/2)$, $b\exp(i\beta\eta)\exp(-i\beta^2\xi/2)$, where $a,b$ are wave amplitudes (taken here to be real) and $\pm\alpha,\beta$ are propagation angles. We also assume that the third beam is weak, e.g. $b < a$. Further we fix $a = 0.5$, $\alpha = 4$, and vary $b$ and $\beta$. As one can clearly see, the two waves with angles $\pm\alpha$ create a static harmonic lattice whose profile is invariant with $\xi$, while the weak third wave component that propagates at an angle $\beta$ makes the lattice imbalanced, so that it carries a net momentum. The index profile of the dynamical lattice arising from the interference of these three waves is described by the function $R(\eta,\xi) = 4a^2 \cos^2(\alpha\eta) + b^2 + 4ab\cos(\alpha\eta)\cos[\beta\eta + (\alpha^2 - \beta^2)\xi/2]$. The structure of this lattice is shown in Fig. 1. The lattice modulation depth and spatial frequency (along $\xi$) grow as the amplitude $b$ and angle $\beta$ of the third plane wave increase. In addition, equation (1) allows the total power or energy flow $U = \int_{-\infty}^{\infty} |q|^2 \, d\eta$ to be conserved.

When $b = 0$ the nonlinear lattice is stationary (has zero momentum) and supports soliton solutions that conserve their profiles upon propagation along the lattice channels. The discrete soliton solution can be obtained by assuming the form $q(\eta,\xi) = w(\eta)\exp(i\lambda\xi)$, where $w(\eta)$ is a real function and $\lambda$ is a propagation constant. A detailed study of the soliton properties in photorefractive stationary lattices can be found in Refs. [9-11]. Here we only recall the fact that there exist an upper $\lambda_{\text{upp}} = E$ and a lower $\lambda_{\text{low}}$ cutoff for the soliton existence, where the lower cutoff $\lambda_{\text{low}}$ is defined solely by the amplitude $a$ and the angles $\pm\alpha$ of the lattice-creating plane waves (Fig. 2). Close to their cutoffs the lattice solitons broaden drastically and cover many transverse lattice periods. The energy flow is a monotonically growing function of the propagation constant. Further we interested in the propagation of relatively broad solitons with energy flow $U \sim 1$ in dynamical lattices with $b \neq 0$. The width of such solitons decreases with increase of $U$. Such solitons (found by solving Eq. (1) at $b = 0$) experience almost no change in their shape when they are launched parallel to the $\xi$ axis (into the distorted lattice) as long as $b < a$. However the soliton center

$$\delta\eta(\xi) = \frac{1}{U} \int_{-\infty}^{\infty} \eta |q|^2 \, d\eta \qquad (2)$$

experiences a progressive shift in the $\eta$ coordinate along the propagation direction of the third plane wave, as shown in Fig. 2(c). Even though the trajectory of the soliton center is rather complicated, it is clear from Fig. 2(c) that the accumulated transverse shift is proportional to the propagation distance, and hence it is possible to introduce an average soliton drift angle as $\phi = \delta\eta/\xi$. In what follows we will show that the soliton drift angle can be controlled by the amplitude $b$ and the angle $\beta$ of third plane wave.

To understand the physical origin of soliton dragging induced by the dynamical optical lattices one can consider a simplified nearly-integrable version of Eq. (1) that can be obtained under the assumptions $S|q|^2, R(\eta,\xi) \ll 1$ and $E \gg 1$. After some rescaling Eq. (1) can be transformed into the cubic nonlinear Schrödinger equation (also encountered in the description of matter waves in optical lattices):

$$i\frac{\partial q}{\partial \xi} = -\frac{1}{2}\frac{\partial^2 q}{\partial \eta^2} - q|q|^2 - R(\eta,\xi)q, \qquad (3)$$

If under appropriate conditions the last term in Eq. (3) can be considered small, then it is possible to use perturbation theory (based on the inverse scattering transform) in order to analyze the evolution parameters of the soliton solutions of the unperturbed Eq. (3). Under these conditions, the tilt $\phi$ (or instantaneous propagation angle) of the soliton $q(\eta,\xi) = \chi \mathrm{sech}[\chi(\eta-\phi\xi)]\exp[i\phi\eta - i(\chi^2 - \phi^2)\xi/2]$ of Eq. (3) changes upon propagation according to

$$\frac{d\phi}{d\xi} = \chi^2 \int_{-\infty}^{\infty} R(\eta,\xi)\mathrm{sech}^2(\chi\eta)\tanh(\chi\eta)d\eta. \qquad (4)$$

Here $\chi$ is the soliton amplitude and the energy flow of this soliton state is given by $U = 2\chi$. Substitution of the lattice profile into Eq. (4) leads to the following expression for the instantaneous propagation angle of the soliton beam when is initially launched parallel to $\xi$ axis, i.e. at $\phi|_{\xi=0} = 0$:

$$\phi = \frac{2\pi ab}{\chi}\left(\frac{\beta-\alpha}{(\beta+\alpha)\sinh[\pi(\beta-\alpha)/2\chi]} - \frac{\beta+\alpha}{(\beta-\alpha)\sinh[\pi(\beta+\alpha)/2\chi]}\right)\left(1 - \cos\frac{\beta^2-\alpha^2}{2}\xi\right). \qquad (5)$$

This expression is valid as long as the conditions $\beta - \alpha \ll \alpha, \beta$, $\beta + \alpha \gg 1$, and $\alpha \neq \beta$, that justify the applicability of perturbation theory for Eq. (1), hold. It is under these conditions when the expression $q(\eta,\xi) = \chi \,\text{sech}[\chi(\eta - \phi\xi)]\exp[i\phi\eta - i(\chi^2 - \phi^2)\xi/2]$ adequately describes soliton profile in the dynamical lattice. Equation (5) clearly shows that it is the last "interference" term in the expression in the lattice profile $R(\eta,\xi)$ that leads to the soliton drift along the positive $\eta$ direction. According to Eq. (5) the propagation angle $\phi$ oscillates periodically between zero and its maximum value, and this explains why the soliton center oscillates as shown in Fig. 2(c). The average drift angle with respect to the $\xi$ axis can be obtained from Eq. (5) by omitting the term $\cos[(\beta^2 - \alpha^2)\xi/2]$ that oscillates upon propagation. According to Eq. (5) the average drift angle grows linearly with the amplitudes $a,b$ of the lattice-creating plane waves, and monotonically increases with the soliton amplitude $\chi$ (or, equivalently, its energy flow). Finally it decreases as the difference $\beta - \alpha$ between plane wave propagation angles increases. This is because the momentum exchange becomes less effective as the lattice "breaths" faster during propagation.

We have verified the predictions obtained on the basis of the reduced model Eq. (3) with direct numerical simulations of Eq. (1) that describes optical beam propagation in actual photorefractive crystals such as SBN.[9,10] The dimensionless parameters that we use below correspond to an input beam with width $\sim 10\,\mu\text{m}$ at the wavelength $0.63\,\mu\text{m}$, launched into a SBN crystal with effective electro-optic coefficient $r_{\text{eff}} = 1.8 \times 10^{-10}$ m/V and biased with a static electric field $\sim 10^6$ V/m. For input, we have used the exact soliton solutions of Eq. (1) when the third plane wave is zero ($b = 0$). By doing so, both the soliton center trajectory as well as the average drift angle was monitored. These results are summarized in Fig. 3 and are in full agreement with the theoretical predictions. One can see from Fig. 3(a) that the soliton drift angle decreases monotonically as the difference $\beta - \alpha$ increases. This should have been anticipated since the soliton does not "feel" the rapid oscillations of the dynamic lattice when $\beta - \alpha$ is large. Notice that at $\beta = \alpha$ the lattice becomes stationary, thus unable to drag the solitons. Consequently the soliton drift angle drops off as $\beta \to \alpha$; this occurs in the region of very small angle differences $\beta - \alpha$, not even visible in Fig. 3(a). The soliton propagation trajectory may depart considerably from the trajectory of a linear beam for such small angle differences.

It is worth noticing that the dragging effect of dynamical lattice is more pronounced for narrow solitons that have higher energy flows (Fig. 3(b)). This effect, that may be surprising at a first glance, is fully consistent with properties of solitons in lattices.[1-8] Narrow solitons experience stronger interactions with stationary lattices, hence their mobility is restricted. However, in our case it is the dynamical lattice that drags the otherwise immobile solitons; therefore stronger interactions experienced by narrow solitons result in more effective momentum exchange, hence in more effective dragging. Finally, as expected, the drift angle varies linearly with $b$ (Fig. 3(c)). Notice that the trajectory of the soliton center can depart considerably from a linear one also for narrow solitons occupying only a couple of lattice sites and in strongly distorted lattices with $b \sim a$, $\beta - \alpha \sim \beta, \alpha$.

In conclusion, we have predicted that lattice solitons can experience a progressive drift in imbalanced optical lattices that exhibit transverse momentum. The soliton drift is caused by the dragging induced by the dynamically-evolving lattice. The drift angle can be effectively controlled by varying the amplitude and the angle of a weak control plane wave. The effect reported here may have promising applications in all-optical steering applications.

*On leave from Physics Department, M. V. Lomonosov Moscow State University, 119899, Vorobiovy Gory, Moscow, Russia. This work has been partially supported by the Government of Spain through BFM2002-2861 and by the Ramon-y-Cajal Program.

# References with titles

# References without titles

## Figure captions

Figure 1. Refractive index profiles of distorted lattices with $\beta = 4.192$ (left) and $\beta = 4.284$ (right) at $b = 0.3$.

Figure 2. (a) Soliton profiles supported by undistorted lattice at $E = 12$, $S = 0.2$. (b) Soliton energy flow versus propagation constant at $E = 12$. (c) Trajectory of integral center of soliton with $U = 1$ in distorted lattice at $\beta = 4.1$, $E = 12$, $S = 0.2$.

Figure 3. (a) Soliton drift angle versus propagation angle of lattice-distorting plane wave at $U = 1$, $b = 0.2$. (b) Soliton drift angle versus energy flow at $\beta = 4.5$, $b = 0.2$. (c) Soliton drift angle versus amplitude of lattice-distorting plane wave at $U = 1$, $\beta = 4.5$. Parameters $E = 12$, $S = 0.2$.

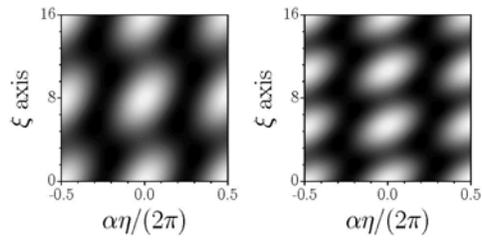

Figure 1. Refractive index profiles of distorted lattices with $\beta = 4.192$ (left) and $\beta = 4.284$ (right) at $b = 0.3$.

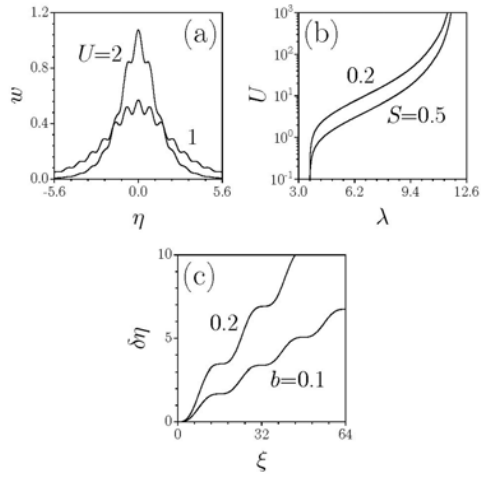

Figure 2. (a) Soliton profiles supported by undistorted lattice at $E = 12$, $S = 0.2$. (b) Soliton energy flow versus propagation constant at $E = 12$. (c) Trajectory of integral center of soliton with $U = 1$ in distorted lattice at $\beta = 4.1$, $E = 12$, $S = 0.2$.

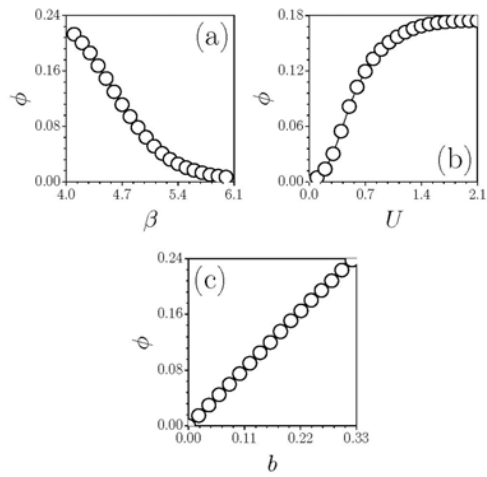

Figure 3.  (a) Soliton drift angle versus propagation angle of lattice-distorting plane wave at $U=1$, $b=0.2$. (b) Soliton drift angle versus energy flow at $\beta=4.5$, $b=0.2$. (c) Soliton drift angle versus amplitude of lattice-distorting plane wave at $U=1$, $\beta=4.5$. Parameters $E=12$, $S=0.2$.